# Mapping social media attention in Microbiology: Identifying main topics and actors

*Nicolas Robinson-Garcia[1,2], Wenceslao Arroyo-Machado[3] and Daniel Torres-Salinas[3,4]*

[1] INGENIO (CSIC-UPV), Universitat Politècnica de València, Valencia, Spain
[2] School of Public Policy, Georgia Institute of Technology, Atlanta, United States
[3] Vicerrectorado de Investigación y Transferencia, Universidad de Granada, Granada, Spain
[4] EC3metrics spin off, Granada, Spain

**One sentence summary:** The aim of this paper is to map and identify topics of interest within the field of Microbiology and identify the main sources driving such attention.

## Abstract

This paper aims to map and identify topics of interest within the field of Microbiology and identify the main sources driving such attention. We combine data from Web of Science and Altmetric.com, a platform which retrieves mentions to scientific literature from social media and other non-academic communication outlets. We focus on the dissemination of microbial publications in Twitter, news media and policy briefs. A two-mode network of social accounts shows distinctive areas of activity. We identify a cluster of papers mentioned solely by regional news media. A central area of the network is formed by papers discussed by the three outlets. A large portion of the network is driven by Twitter activity. When analyzing top actors contributing to such network, we observe that more than half of the Twitter accounts are bots, mentioning 32% of the documents in our dataset. Within news media outlets, there is a predominance of popular science outlets. With regard to policy briefs, both international and national bodies are represented. Finally, our topic analysis shows that the thematic focus of papers mentioned varies by outlet. While news media cover the wider range of topics, policy briefs are focused on translational medicine, and bacterial outbreaks.

**Keywords:** altmetrics; science mapping; Twitter; news media; policy documents; microbiology

# Introduction

In a rapidly changing scholarly communication system, the number of publications grows exponentially (Van Noorden, Maher and Nuzzo 2014), increasing researchers' difficulties to tap into the relevant literature and identify topics of interest and research fronts (Redfern, Cobo and Herrera-Viedma 2018). In this context, science mapping solutions can become key tools for easing researchers' burden. In this study we aim at identifying topics of social interest within the field of Microbiology and exploring the mechanisms which might explain such social interest. We do so by using data extracted from mentions from news media, policy documents and Twitter to scientific publications, instead of citation data, as it has been traditionally been conducted. Despite the expansion of the use of bibliometric techniques and methods to analyze specific scientific fields and areas, they have been rarely applied to the field microbiology (Nai 2017). The ones that have been conducted have either focused on a particular topic or aspect (Brandt *et al.* 2014) or have focused on the main actors and regions active in the field and their evolution over time (Vergidis *et al.* 2005). But, to our knowledge, no study has tried to fill the science-society gap, by aiming at connecting research topics with societal interest.

Science mapping has been extensively used in the context of research evaluation for identifying research priorities (Cassi *et al.* 2017), to aid governance of specific areas (Rotolo *et al.* 2014) or to profile institutions' research portfolio (Rafols, Porter and Leydesdorff 2010). The expansion of science mapping applications is largely derived to the free availability of academic software and tools that are constantly maintained and updated (Cobo *et al.* 2011). These maps usually combine publication data with citation data, although there are notable exceptions (Klavans and Boyack 2014). In this paper, we use altmetric data in combination with publications.

Altmetrics have become a promising research front in the field of research evaluation and scholarly communication. They are based on the notion that non-formal channels of scholarly communication are shifting to the Internet (Priem 2014). Therefore, by tracking these alternative channels, it is possible to identify and access literature which might not only be relevant to scientists, but also to lay people. Although most interest on altmetrics has focused on applying it for research assessment (Robinson-Garcia, van Leeuwen and Rafols 2018), they can also be used as tools for discovery. In this sense, altmetric data has not been used for science mapping until quite recently (Didegah and Thelwall 2018) pointing out towards it interest as a descriptive tool to showcase thematic landscapes (Wouters, Zahedi and Costas 2018).

## Objectives

The main goal of this paper is to visualize the main topics of social interest as identified via altmetric data. Moreover, we will explore how topics are captured by social media and which are the main drivers of such attention. For this, we focus on three specific altmetric sources: Twitter, news media and policy documents. The selection of these sources is due to the following reasons. Twitter is the main general platform feeding altmetric data both in coverage of publications as well as intensity (Robinson-García *et al.* 2014) and is the most researched of the social media platforms conforming altmetrics (Thelwall *et al.* 2013; Robinson-Garcia *et al.* 2017). Policy mentions and news media are the most robust sources in the sense that they are theoretically easier to interpret, and hence, to understand the underlying meaning behind them.

# Materials and Methods

Here we analyze Twitter, news media and policy briefs' mentions to publications in the field of Microbiology. In 22 October 2018, we retrieved a total of 382,998 records of publications indexed in the subject categories of Microbiology and Biotechnology & Applied Microbiology from Web of Science in the 2012-2018 period. Altmetric data was obtained from

Altmetric.com, one of the main secondary providers of altmetric data (Robinson-García *et al.* 2014). Prior altmetric data is scarce and incomplete as this database started to systematically retrieve altmetric data in 2011. To query Altmetric.com we need to use publications' Digital Object Identifiers (DOI), this means that all publications without DOI will be lost from our final dataset. 88.2% of the dataset included DOI numbers. After querying Altmetric.com we identified a total of 174,799 distinct publications which are at least mentioned once by any of the sources covered by Altmetric.com. Furthermore, we downloaded all mentions retrieved from Altmetric.com, that is, a total of 1,594,856 records. These records do not only indicate the publication being mentioned, but more importantly, the author of such mention. Table 1 includes some descriptive of the total number of mentions by platform, and papers mentioned. As observed, around 90% of mentioned papers were mentioned by Twitter users, being the most prominent source of altmetrics. News stories cover around 10% of mentioned publications, while policy documents barely cite 2% of mentioned publications.

Two mapping approaches were followed in this study. First, based on the dataset of mentions, we conducted a two-mode network analysis to identify the most influential actors and the papers being mentioned by them. That is, papers are connected to each either through the actors which mention them. These actors can be Twitter users, news media or organizations publishing policy briefs. Two-mode networks consist on two types of actors (i.e., publications and altmetric sources: Twitter accounts, organizations publishing policy briefs and news media) connected by a direct relation between each other (i.e., altmetric source mentions publication).

**Table 1.** Descriptive of mentions and papers mentioned in Altmetric.com by platform from all publications indexed in Web of Science subject categories Microbiology and Biotechnology & Applied Microbiology for the 2012-2018 period. In bold platforms used in this paper.

| Platforms | Mentions | Share of mentions | Number of mentioned papers | Share of mentioned papers |
|---|---|---|---|---|
| **Tweet** | **1345909** | **84.4** | **156912** | **89.8** |
| **News story** | **80485** | **5.1** | **16529** | **9.5** |
| Facebook post | 73189 | 4.6 | 28394 | 16.2 |
| Blog post | 29622 | 1.9 | 18090 | 10.4 |
| Patent | 24001 | 1.5 | 10100 | 5.8 |
| Google+ post | 14834 | 0.9 | 6716 | 3.8 |
| Wikipedia page | 10243 | 0.6 | 7623 | 4.4 |
| **Policy document** | **4414** | **0.3** | **3295** | **1.9** |
| F1000 post | 4175 | 0.3 | 3589 | 2.1 |
| Reddit post | 3769 | 0.2 | 3120 | 1.8 |
| Peer review | 1767 | 0.1 | 751 | 0.4 |
| Weibo post | 1083 | 0.1 | 298 | 0.2 |
| Video | 1040 | 0.1 | 799 | 0.5 |
| Q&A post | 224 | 0.0 | 209 | 0.1 |
| Pin | 53 | 0.0 | 46 | 0.0 |
| LinkedIn post | 48 | 0.0 | 48 | 0.0 |
| *Total* | *1594856* | *100.0* | *174799* | *100.0* |

In a second step, we aim at identifying the most discussed topics by each media analyzed. In this case, we create a thematic landscape based on terms contained in the titles of papers with mentions in Altmetric.com. This landscape offers an overview of what is being discussed in social media but does not provide any information about the intensity of such discussions. For this, we overlay the number of mentions from each of the selected platform on the thematic landscape. This term map uses binary counting, that is, we do not consider how many times terms occur in a single title but the number of times they occur in different titles from other publications.

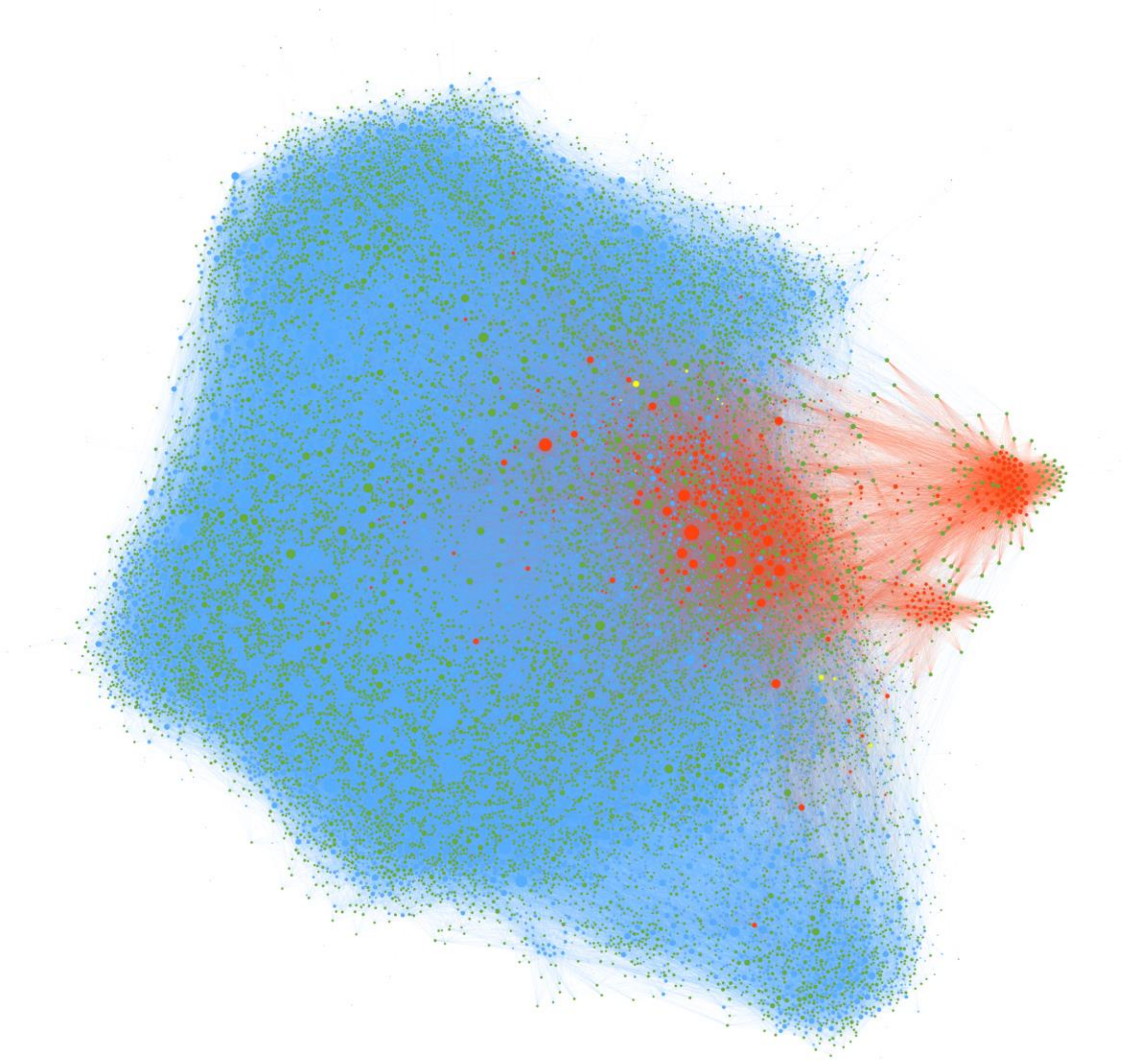

**Figure 1.** Two-mode network of Microbiology publications and altmetric actors mentioning them. We only show 6% of the network for displaying reasons. Publications are represented in green and they correspond to all publications included in Altmetric.com from the Web of Science subject categories of Microbiology and Biotechnology & Applied Microbiology. Nodes in blue are Twitter accounts, in red are news media and in yellow, organizations producing policy briefs. Map created with Gephi v. 0.9.1

# Results

Figure 1 shows a two-mode network of publications and the actor mentioning such publications. The largest portion of mentions to literature in the field of Microbiology come from Twitter discussions. Twitter users represent around 50% of the nodes of the network and involve 94% of the links observed. While some of the discussion generated from Twitter revolves around stories published by news media, there is a large portion of literature which is only discussed in Twitter. Literature discussed solely in Twitter are literature reviews, publications related with crystallography or news stories affecting either academia (e.g., open science) or specifically with regard to the field of microbiology ( e.g., calls for best practices).

There is a smaller cluster of news media mentions to literature which are not discussed in Twitter. These news stories come from local and regional US news media such as Mississippi News Now or KSLA News (from Los Angeles), as well as agencies (e.g., EuropaPress). In the case of news media mentions to publications also discussed in Twitter, we find internationally renowned media such as The Economist, EurekaAlert! or Scientific American. Most discussed publications in this cluster have to do with chemotherapy solutions and topics related with oncology or advancements on the development of vaccines to prevent viral diseases. On the contrary, publications discussed also in Twitter are related with topics which seem to be less applied and more appealing to the curious mind. Here we find papers on the identification of new viruses, calls for the preservation of microbial diversity or new brewing techniques for beer production. Also, news media outlets vary, although there is a high degree of US regional and local media, also some national news media are present such as PBR. However, in all cases we must note high predominance of US media. Policy briefs are scarce and tend to connect publications which are both, discussed in Twitter and by news media. These tend to cite publications discussing specific health issues such as outbreaks in animals or humans in different places of the world. Also, some of the studies cited target specific human groups such as pregnant women.

There is a total of 216735 Twitter accounts that mention at least one publication. 66.05 percent of Twitter accounts mention only one publication, while 12.37 percent do so with 2, 5.42 percent with 3 and from 7 mentions the percentage is below 1 percent. The mean of unique mentions is 5.44 (± 53.02). In figure 2 we focus of the top 25 accounts driving the conversation in Twitter. We distinguish between the number of tweets mentioning publications and the actual number of papers which are mentioned. We manually assign an account type to these 25 cases. In all of them, although numbers are similar, there are always papers which are mentioned in more than one occasion. Even more, there is one account (@FarmFairyCrafts) which has only mentioned 5 publications, but these have been mentioned in more than 3500 tweets. While in this case, the account belongs to a firm in Texas, we observe that 12 of these top 25 accounts are bots, followed by 7 accounts from academics, related to scientific journals, one to a news media and one to a physician. These 12 bots are responsible for 4% of the tweets which are directed to 32% of the papers in our sample.

| | ALTMETRIC DATA | | ACCOUNT DATA | | | |
|---|---|---|---|---|---|---|
| **Account name** | **Nr Papers tweeted** | **Nr Mentions** | **Account type** | **Nr Tweets** | **Following** | **Followers** |
| @AntibioticResis | 6497 | 7699 | Bot | 27300 | 931 | 9024 |
| @yeast_papers | 6147 | 6342 | Bot | 33300 | 3 | 1349 |
| @rnomics | 3863 | 6218 | Bot | 16200 | 119 | 2295 |
| @jcamthrash | 5441 | 5689 | Academic | 19900 | 489 | 7008 |
| @FrontMicrobiol | 5007 | 5290 | Journal | 7222 | 816 | 5913 |
| @EvolvedBiofilm | 3898 | 4940 | Academic | 29300 | 1156 | 4055 |
| @msmjetten | 2948 | 4929 | Academic | 35800 | 507 | 2719 |
| @micro_papers | 4448 | 4587 | Bot | 12200 | 11 | 210 |
| @MicrobiomePaper | 4218 | 4465 | Bot | 19800 | 53 | 3154 |
| @biofilmPapers | 4305 | 4416 | Bot | 14000 | 64 | 920 |
| @pseudo_papers | 4030 | 4048 | Bot | 16400 | 35 | 797 |
| @ndm1bacteria | 3221 | 3981 | Press | 15500 | 49 | 581 |
| @PLOSPathogens | 2610 | 3978 | Journal | 6601 | 2530 | 21700 |
| @Immunol_papers | 2920 | 3969 | Bot | 61700 | 0 | 771 |
| @animesh1977 | 3129 | 3962 | Professional | 954 | 1059 | 1592 |
| @phy_papers | 3906 | 3946 | Bot | 20800 | 1 | 2252 |
| @custom_ms | 3777 | 3865 | Bot | 11100 | 2 | 27 |
| @BIOCIENCIA2013 | 3192 | 3717 | Academic | 67200 | 565 | 967 |
| @MicrobiomDigest | 3356 | 3597 | Academic | 34000 | 15100 | 20400 |
| @bmgphd | 2636 | 3589 | Academic | 8375 | 228 | 496 |
| @FarmFairyCrafts | 5 | 3578 | Company | 755 | 20700 | 27500 |
| @ASMicrobiology | 2895 | 3251 | Academic | 18900 | 218 | 37400 |
| @BioinformaticsP | 3052 | 3062 | Bot | 4865 | 27 | 318 |
| @NatureRevMicro | 2686 | 3059 | Journal | 10300 | 1340 | 38800 |
| @transcriptomes | 2638 | 2804 | Bot | 21300 | 4 | 897 |

**Figure 2.** Top 25 Twitter accounts mentioning publications included in Altmetric.com from the Web of Science subject categories of Microbiology and Biotechnology & Applied Microbiology.

Mentions to publications from news media seem to be ridden by news agencies and media specialized in scientific literature. In figure 3A we observe the top 15 news media mentioning microbial literature. EurekaAlert!, a service that provides news releases to journalists, stands out as the main news media. The rest of the list is mostly populated by media focused on medical literature (e.g., Health Medicinet, MedicalXpress).

In the case of policy briefs (Figure 3B), the World Health Organization is the most prominent institution citing microbial literature. Along with this organization, we find other international bodies such as the Food and Agriculture Organization of the United Nations (FAO), but most of the top institutions citing microbial literature in their policy briefs are of a national or regional scope. Here we highlight the presence of the UK Government, the US Centers for Disease Control and Prevention, the European Union or the Netherlands Environmental Agency (PBL).

Next, we expand and focus on the topics discussed by each social source. For this, figure 4 maps terms included in titles of all microbial publications indexed in Altmetric.com and overlays the focus of discussion for each source. It shows the base map where nodes represent words and noun phrases from titles and colors represent clusters of topics. Seven large topics are identified. The red clusters relates with molecular and cell biology. The yellow cluster represents papers closer to clinical and translational medicine related to virus biology and immunity. The blue cluster in the bottom right, refers to bacterial infections and hubs. A light blue cluster is spread in the middle of the map on the left side of the red cluster and on the left side of the blue one. Such spread is due to the fact that it refers to methodological approaches and techniques for discovery. A separate orange cluster can be observed on the bottom left. Here we find terms related with taxonomies of bacterial species. The large green cluster on the left side represents bioengineering research. Lastly, we observe a purple cluster in the middle of the map just beneath the red cluster. Here we observe terms such as progress, current status or future, which point at future prospects and state of the art papers.

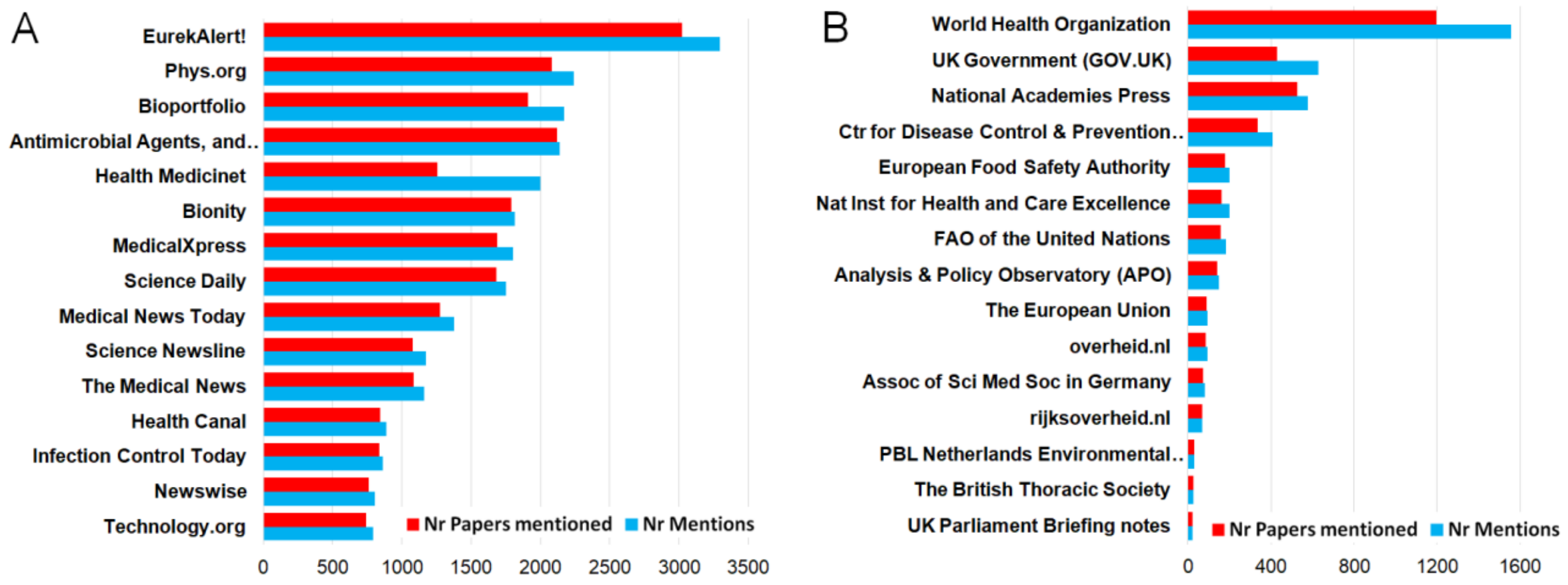

**Figure 3.** Top 15 **A** news media and **B** international organizations mentioning publications included in Altmetric.com from the Web of Science subject categories of Microbiology and Biotechnology & Applied Microbiology.

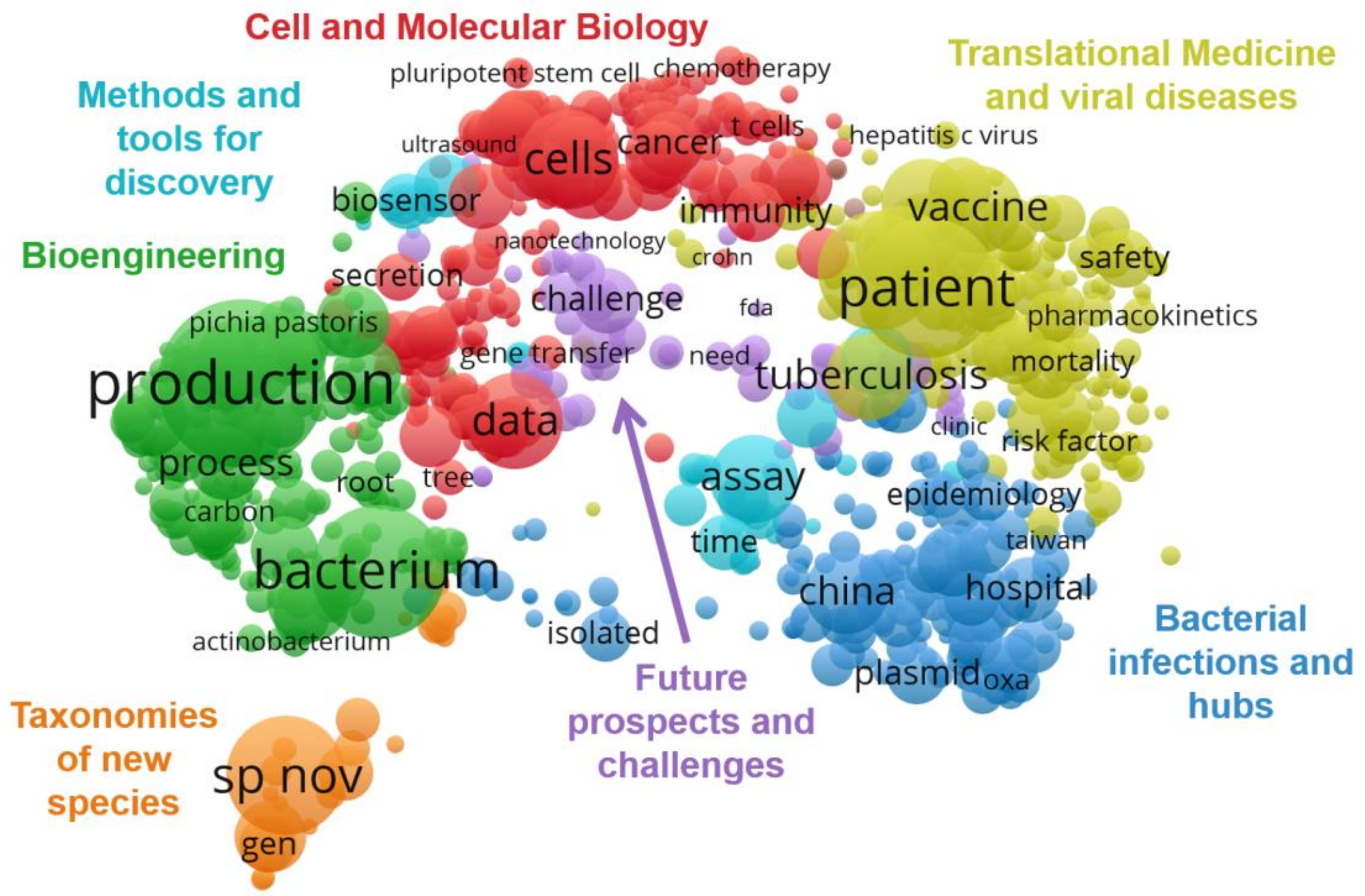

**Figure 4.** Top 60% most relevant terms occurring in titles of publications included in Altmetric.com from the Web of Science subject categories of Microbiology and Biotechnology & Applied Microbiology. Minimum threshold: 50 co-occurrences. Clusters of topics using a modularity-based community detection algorithm (Waltman and Eck 2013). Map created using VOSviewer v. 1.6.10

Twitter mentions are distributed among translational medicine, and future prospects and challenges, as well as bacterial outbreaks (Figure 5A). News media covers a broader scope of topics. We observe a higher prevalence in those related with translational medicine and bacterial outbreaks and infections, but there are also mentions to bioengineering and molecular and cell biology (Figure 5B). Policy mentions are more thematically constrained, focusing mainly on bacterial infections and outbreaks as well as on translational medicine, with almost no mention to any other topic within the field (Figure 5C).

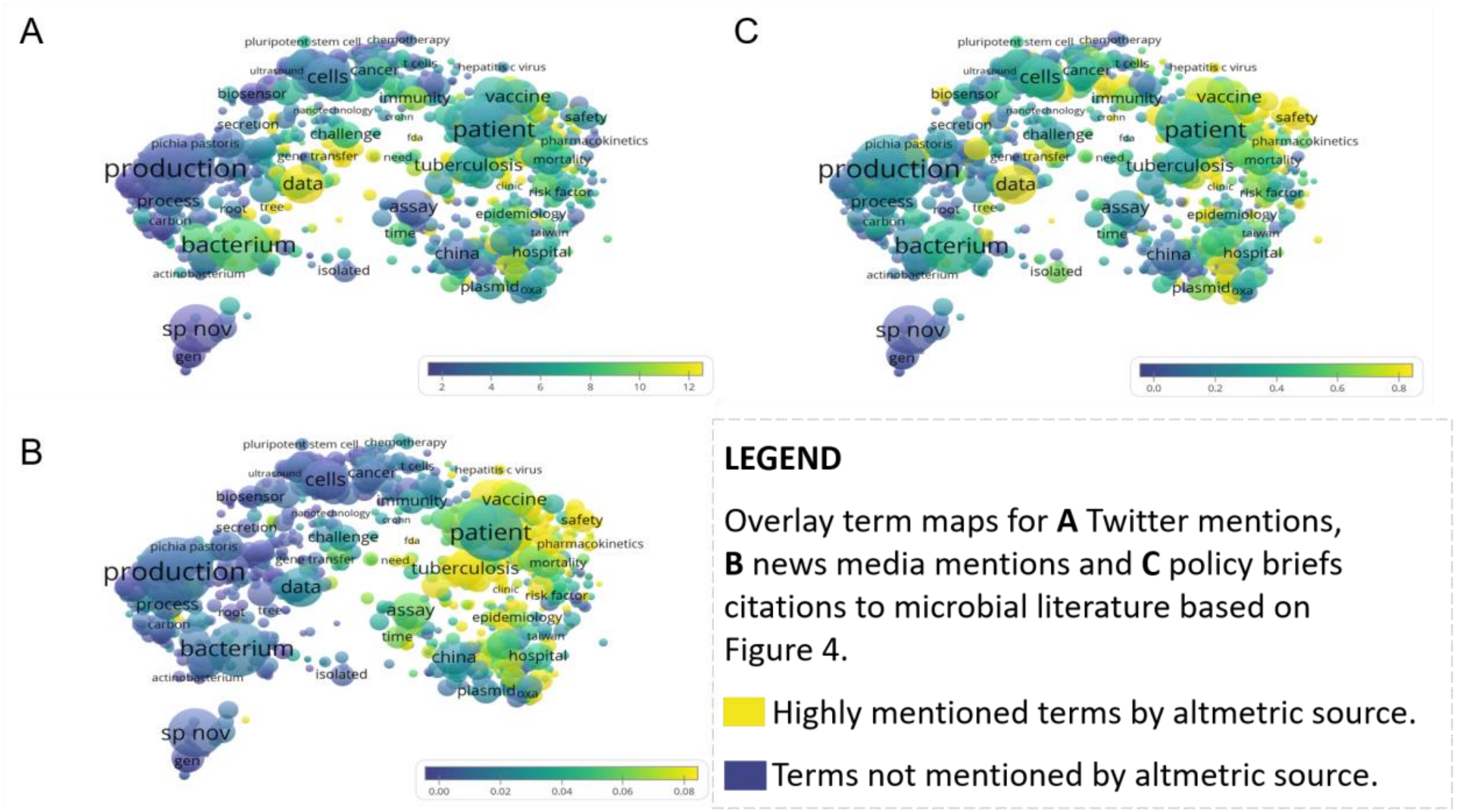

**Figure 5.** Overlays maps based on figure 4. **A** overlays tweet mentions to terms, **B** overlays news media mentions and **C** overlays mentions from policy briefs. Map created using VOSviewer v. 1.6.10

# Discussion

This paper analyzes and identifies topics of social discussion on microbial literature by analyzing mentions to scientific publications in Twitter, news media and policy briefs. We identify and describe which are the actors or channels riding such discussion. For this we make use of mapping techniques and network analysis. Not surprisingly, most of the mentions identified come from Twitter activity. Interestingly, we do find separate clusters of discussion (figure 1): a large cluster of tweets with half of it closely related to news media and two isolated clusters of news media mentioning papers. This signifies that there are publications which drive news media attention but are not discussed socially via Twitter, while there are many other papers which generate a large amount of Twitter attention but are not of interest neither for news media or policy briefs. Policy briefs tend to cite publications which also drive news media and Twitter attention. Differences on news media attention seem to revolve around the locality or globality of topics. Altmetric.com's selection of news media is severely biased towards English language (Robinson-García *et al.* 2014), which explains why news media in these isolated clusters are mainly local and regional US media. Regarding thematic differences, we note that most of the papers mentioned solely by Twitter seem to revolve around academic discussion, confirming the role of Twitter as a non-formal channel of communication for academics, rather than for lay people (Sugimoto *et al.* 2017). On the other hand, we observe that papers mentioned by news media, policy briefs and Twitter combined, are related with social issues such as advancements on therapies and vaccines and the identification of viral or bacterial diseases.

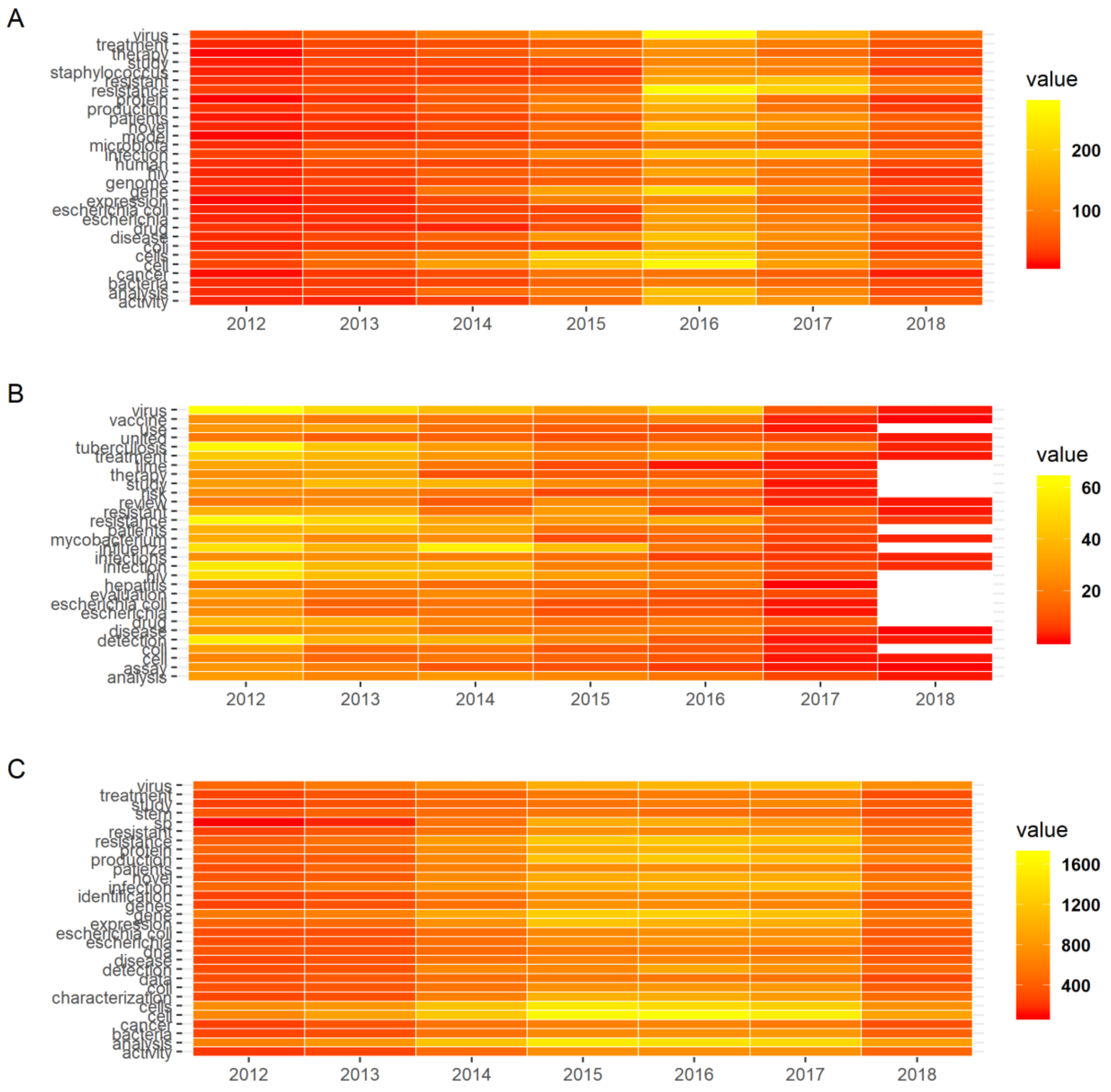

**Figure 6** Heatmap of the top 30 most occurring noun phrases in papers mentioned by **A** news media, **B** policy briefs and **C** Twitter accounts.

Our analysis on the top Twitter accounts mentioning microbial literature (Figure 2) confirm that the existence of bots compromises to a great extent any use of tweet mentions as a means to assess the societal relevance of specific papers (Haustein *et al.* 2016; Robinson-Garcia *et al.* 2017). In the case of news media (Figure 3A), there is a preponderance of science specialized media within the media citing the most microbial literature, as well as the above mentioned biased towards English language which obscures local interest from non-English speaking countries, limiting to a great extent the identification of socially interesting topics in peripheral regions (Alperin 2013). Organization publishing policy briefs which cite microbial literature are both of an international and national scope (Figure 3B). Furthermore, while some of them are field specific (e.g., the British Thoracic Society), others are of a much broader breadth and social influence (e.g., Food and Agriculture Organization of the United Nations or the European Union).

Finally, we observe differences on the topics discussed by each of these sources (Figure 4). While news media seem to cover a wider range of topics, Twitter mentions seem to be more related to future prospects of the field as well as translational research and virology. Policy briefs, on the other hand, are thematically focused on bacterial infections and hubs, and viral diseases and translational medicine. This reveals a great thematic dependency on what drives more social attention (Noyons and Rafols 2018) which compromises general statements and suggestions to push all scientific efforts towards socially relevant topics (e.g., UK's Research Excellence Framework), as it would work on the benefit of some areas and on the detriment of others.

In this paper we have shown that the combination of advanced visualization techniques, network analysis and different altmetric data sources, provides valuable information not only to identify topics of social interest, but also to better assess how such attention is generated and better interpret such differences on topics and communities. While these analyses are still rare with most of the efforts analyzing altmetrics focused on research assessment (Bornmann 2014), recent calls for the use of altmetrics for contextualizing how social attention of research is generated and identifying areas of social engagement (Robinson-Garcia, van Leeuwen and Rafols 2018; Wouters, Zahedi and Costas 2018) will hopefully help to develop advanced methods which can better inform academics and research managers to spot and understand how social attention is generated.

With regard to the topic clusters identified generating more attention by altmetric source, while the results are somehow expected (e.g., viral diseases being of higher interest in news media than bioengineering), they allow to validate the combination of methods. These methods can be of greater interest if more closely refined and directed at specific targets (e.g., social interest of Zika in Latin America) to better understand how research outputs are perceived and used by the public. For instance, by targeting specific terms or noun phrases and analyzing frequency of occurring in publications mentioned by an altmetric data sources and monitoring peaks of attention, similarly to what we show in Figure 6.

## Acknowledgements

The authors are grateful to Lydia Robinson-Garcia from the CeMM Research Center for Molecular Medicine of the Austrian Academy of Sciences for assessing on the description and interpretation of the term maps. Nicolas Robinson-Garcia is currently supported by a Juan de la Cierva-Incorporación grant from the Spanish Ministry of Science, Innovation and Universities.